\documentclass[prl,aps,amssymb,psfig,twocolumn,showpacs]{revtex4}

\usepackage{graphicx}

\begin{document}

\title{Magnetization transport and quantized spin conductance} 
\author{Florian Meier and Daniel Loss}
\affiliation{Department of Physics and Astronomy, University of Basel, 
Klingelbergstrasse 82, 4056 Basel, Switzerland}
\date{\today}

\begin{abstract}
We analyze  transport of magnetization in insulating systems described by a 
spin Hamiltonian. 
The magnetization current through a quasi-one-dimensional magnetic 
wire of finite length suspended between two bulk magnets is determined by
the spin conductance which remains finite in the ballistic limit due to
contact resistance.
For ferromagnetic systems, magnetization transport can be
viewed as transmission of magnons, and the spin conductance depends on the
temperature $T$. For antiferromagnetic isotropic spin-$1/2$ chains, 
the  spin conductance is quantized in units of order $(g \mu_B)^2/h$ 
at $T=0$. 
Magnetization 
currents produce an electric field and, hence, can be measured directly. For 
magnetization transport in electric fields, phenomena analogous to the Hall 
effect emerge.
\end{abstract}


\maketitle

Transport of magnetization in various magnetic systems has received 
considerable attention both theoretically and 
experimentally~\cite{wolf:01,awschalom:02,forster:75,slonczewski:89}.
A spatially varying magnetic field
gives rise to a current of magnetic dipoles~\cite{forster:75,slonczewski:89}, 
similar to the transport of electric charge driven by an electric 
gradient. Here we consider insulating magnets
described by a spin Hamiltonian, where magnetization can
be transported by excitations such as magnons and spinons without
transport of charge. 
Theoretical work on such systems has been focused on the long-wavelength 
limit for magnets with translational invariance~\cite{forster:75,castella:95}.
  
In contrast, we propose to investigate magnetization transport in systems
with broken translational invariance. In particular, we consider a 
quasi-one-dimensional system of finite length, e.g., a spin chain 
sandwiched between two bulk magnets which act as reservoirs for magnetization,
where the magnetic field gradient is nonzero only over the system. 
Then, the magnetization current is determined by the 
spin {\it conductance} $G$ which remains finite in the ballistic
limit due to the contact resistance between 
the reservoirs and the system, in analogy to electronic transport in 
mesoscopic
systems~\cite{datta}. This is in stark contrast to the spin conductivity
which diverges in the ballistic limit due to 
translational invariance~\cite{forster:75,castella:95}.  
Here, we derive the spin conductance $G$ for both ferromagnetic (FM) and 
antiferromagnetic (AF) systems. We find that, 
for FM systems, magnetization transport can be
viewed as transmission of magnons and the conductance is temperature 
dependent. For the AF spin-$1/2$
chain, the conductance has a value of order $(g \mu_B)^2/h$, where
$g$ is the gyromagnetic ratio and $\mu_B$ the Bohr magneton. Further, 
spin currents produce 
an electric field which allows one to measure $G$. We discuss 
magnetization transport in an external electric field and show that phenomena 
analogous to the Hall effect exist.

{\it Ferromagnetic systems. --} We first discuss a system
with isotropic FM exchange interaction in a magnetic
field ${\bf B} ({\bf x}_i) = B_i {\bf e}_z$. The spins occupy the sites 
${\bf x}_i$ of a simple $d$-dimensional lattice with lattice constant $a$,
\begin{equation}
\hat{H} = J \sum_{\langle ij \rangle} \hat{\bf s}_i \cdot \hat{\bf s}_j
+ g \mu_B \sum_i B_i \hat{s}_{i,z},
\label{eq:hham}
\end{equation}
with $J<0$. Here, $\hat{\bf s}_i$ is the spin operator of the spin with 
spin quantum number $S$ at ${\bf x}_i$, and $\langle ij\rangle$ denotes nearest
neighbor sites.
For spatially constant $B_i=B>0$, the 
elementary excitations of the system are magnons with dispersion~\cite{mattis}
\begin{equation}
\epsilon_{\bf k} \simeq g \mu_B B + |J| S a^2 k^2
\label{eq:fmagn}
\end{equation}
which carry a magnetic moment $- g \mu_B {\bf e}_z$. Here,
${\bf k}$ is the magnon wave vector. For temperature $T\ll g \mu_B B/k_B$, 
the magnon
density is small and the noninteracting-magnon theory is valid for all $d$. 

\begin{figure}
\centerline{\mbox{\includegraphics[width=8cm]{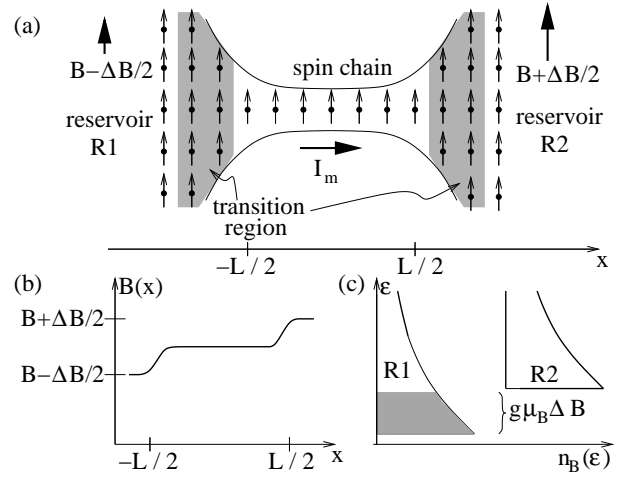}}}
\caption{(a) Proposed experimental setup for the measurement of a
magnetization 
current $I_m$. (b) A magnetic field difference $\Delta B$ between the two 
bulk systems gives rise to $I_m = G \Delta B$. (c) 
$\Delta B$ shifts the Bose functions $n_B (\epsilon)$ in the reservoirs 
R1, R2. Magnons with energies $\epsilon$ within the shaded region in R1 are not
transmitted to R2.
}
\label{Fig1}
\end{figure}

We now consider a setup for a magnetization transport 
measurement as sketched in Fig.~\ref{Fig1}(a). A spin chain extends from 
$x=-L/2$ to $L/2$ and is suspended between two large three-dimensional (3D)
reservoirs, R1 and R2. $L \gg a$ is sufficiently small that magnons
propagate ballistically through the chain. The reservoirs narrow 
adiabatically towards the 
chain [``transition region'' in Fig.~\ref{Fig1}(a)]. The system 
is still described by Eq.~(\ref{eq:hham}), with the sites 
${\bf x}_i$ occupying a bounded region in space [Fig.~\ref{Fig1}(a)]. A small 
spatially varying magnetic field $\delta B(x) {\bf e}_z$ with 
$\delta B(x)  = -\Delta B/2$ ($\Delta B/2$) for $x < -L/2$ ($x> L/2$) is
superimposed on the offset field $B {\bf e}_z$ for $t>0$ [Fig.~\ref{Fig1}(b)].
For $|x|<L/2$, $\delta B(x)$ interpolates smoothly between the 
values $\pm \Delta B/2$ in the reservoirs. 
The field gradient results in a magnetization current $I_m$ from R1 to R2. 
In linear response theory, $I_m$ can be expressed in terms of the spin 
conductivity  $\sigma (x,x^\prime,\omega)$, 
\begin{equation}
I_m (x,\omega) = \int d x^\prime \, \sigma (x,x^\prime,\omega) 
\partial_{x^\prime} \, \delta B (x^\prime, \omega).
\label{eq:def-cond}
\end{equation} 
To calculate $I_m(x,\omega)$, knowledge of $\sigma$  for 
$x, x^\prime \in [-L/2,L/2]$  is sufficient because 
$\partial_{x^\prime} \, \delta B (x^\prime, \omega)=0 $
inside the reservoirs. For a quasi-one-dimensional system, due to the 
continuity equation, 
$\sigma$ is related to the susceptibility $\chi$ 
by $\sigma (q,\omega) =
- i \omega \chi(q,\omega)/q^2$~\cite{remark1}. In the noninteracting-magnon 
approximation,
\begin{equation}
\chi(q,\omega) =  - \frac{(g \mu_B)^2}{\hbar}  
\int_{-\infty}^{\infty} \frac{d k}{2 \pi} \frac{n_B(\epsilon_{k+q})-
n_B(\epsilon_k)}{(\epsilon_{k+q}-\epsilon_k)/\hbar + \omega + i 0 }.
\label{eq:fm-susc}
\end{equation}
Here, $n_B(\epsilon) = 1/[\exp(\beta \epsilon)-1]$ is the Bose distribution 
function and $\beta=1/k_B T$. In the limit $\omega \rightarrow 0$ of
a dc field, from Eq.~(\ref{eq:fm-susc}) we find that
$\lim_{\omega \rightarrow 0}  \sigma (x,x^\prime,\omega) = 
(g \mu_B)^2 n_B (g \mu_B B)/h $ is independent of $x$ and $x^\prime$. 
Integrating over $x^\prime$ in Eq.~(\ref{eq:def-cond}), we find that 
\begin{equation}
I_m (x) = \frac{(g \mu_B)^2}{h} n_B (g \mu_B B) \Delta B = 
G \Delta B
\label{eq:fm-current}
\end{equation}
is constant and depends only on the difference of magnetic fields in the 
reservoirs, 
$\Delta B$. Although magnetization is transported ballistically, the 
spin {\it conductance} $G$ remains finite due to the contact resistance 
for magnons 
between reservoirs and the system, similar to the related phenomenon in
charge transport~\cite{datta}. 

In FM systems, the magnetization current is carried by 
magnons. This allows us to 
reproduce Eq.~(\ref{eq:fm-current}) from the Landauer-B\"uttiker 
approach~\cite{datta}. The field difference $\Delta B$ switched on at 
$t=0$ results in a shift of the magnon energies $\epsilon$ 
in the reservoirs [Eq.~(\ref{eq:fmagn})] and of the magnon distribution 
functions $n_B(\epsilon)$ [Fig.~\ref{Fig1}(c)]. Hence, a nonequlibrium
situation is established. The magnetization in the reservoirs relaxes towards
the new equilibrium values by magnetization transport from R1 to R2, i.e.,
the magnetization current $I_m$. 
All magnons incident on the spin chain from R2 are transmitted into 
R1~\cite{remark2}. In 
contrast, magnons with $\epsilon \in [g \mu_B
(B - \Delta B/2), g \mu_B (B + \Delta B/2)]$ are not transmitted from 
R1 to R2. This results in a net magnetization transport current
\begin{eqnarray}
I_m &=& g \mu_B \int_{0}^{g \mu_B \Delta B} d \epsilon \, 
v(\epsilon) \rho (\epsilon)
n_{\rm B} (\epsilon+ g \mu_B B) 
\nonumber \\
& \simeq &\frac{(g \mu_B)^2}{h} n_B(g \mu_B B) \Delta B = 
G \Delta B, 
\end{eqnarray}
where $v(\epsilon) = \partial_{k_x} \epsilon_{k_x}/\hbar$ is the magnon 
velocity and $\rho (\epsilon) = 1/h v(\epsilon)$ is the magnon density of 
states in the spin chain. 

If the system connecting R1 and R2 consists of several chains
with finite interchain exchange $J^\prime$, 
$G = (g \mu_B)^2 \sum_{k_\perp} n_B (g \mu_B B+ \epsilon_{k_\perp})/h$, 
where $\epsilon_{k_\perp}$ is the energy of the transverse magnon mode. 
At $T=0$, $G=0$ because the system and the reservoirs are in the 
spin-polarized ground state. 

\begin{figure}
\centerline{\mbox{\includegraphics[width=8cm]{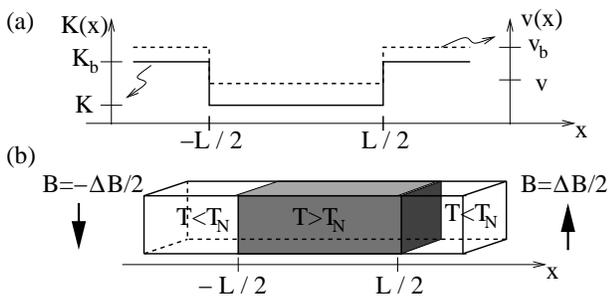}}}
\caption{
(a) The transition 
from the 3D AF ordered bulk to the 
spin-$1/2$  chain is modeled by spatially  varying $K(x)$  (solid line) 
and $v(x)$ (dashed line) in Eq.~(\ref{eq:afm-lagrange}). 
(b) setup for a transport measurement in an AF system.
}
\label{Fig2}
\end{figure}

{\it Antiferromagnetic systems. --} As we show next, magnetization transport 
in antiferromagnets is significantly different from ferromagnets but similar
to charge transport in Fermi liquids.
In an AF chain with half-integer spin,  the elementary  
excitations are massless, and we will show that $G \neq 0$ even  at $T=0$. 
The spin-$1/2$ chain is believed to capture the essential 
features~\cite{fradkin:91,haldane:80,haldane:83,affleck:89}.
Thus, we now consider a spin-$1/2$ chain with isotropic AF 
exchange interaction $J>0$ in Eq.~(\ref{eq:hham})
suspended between two AF reservoirs~\cite{remark3}. 
For $t>0$, a magnetic field $B(x)$ is applied along ${\bf e}_z$ such that
$B(x) = -\Delta B/2$ ($\Delta B/2$) for $x<-L/2$ ($x> L/2$). By a 
Jordan-Wigner
transformation and subsequent bosonization, the spin chain can be mapped onto 
a Luttinger  liquid (spinless fermions). Then, at $T=0$, the Euclidean 
Lagrangian 
flows into a massless free theory under renormalization 
group~\cite{fradkin:91,haldane:80,affleck:89}, 
\begin{equation}
{\mathcal L}_E = \int dx \frac{K}{2} \left[ \frac{1}{v} (\partial_\tau \phi)^2
+ v (\partial_x \phi)^2 \right],
\label{eq:afm-lagrange}
\end{equation}
where $K=2$, $v= (\pi/2) Ja/\hbar$, and the homogeneous part of $\hat{s}_z$ 
is identified with $\partial_x \phi/\sqrt{\pi}$. 
The imaginary-time spin conductivity
is $\sigma (q,\omega_n) = (g \mu_B)^2 (v/\pi \hbar K) 
\omega_n/(\omega_n^2 + v^2 q^2)$~\cite{shankar:90}.
However, in order to calculate $G$, it is not sufficient to evaluate the dc 
limit $\omega \rightarrow 0$ of $\sigma(q,\omega)$ because 
the elementary excitations change on propagation from the reservoirs 
(magnons) through the chain (spinons). 
Following the related analysis for 
charge transport through a Luttinger liquid coupled to Fermi 
leads~\cite{maslov:95}, we model the transition from 
a 3D ordered AF state to the spin chain
by spatially varying $K(x)$ and $v(x)$ in the Lagrangian
Eq.~(\ref{eq:afm-lagrange}). For simplicity, we assume that 
$K(x)$ and $v(x)$ change discontinuously from the values of the spin chain to 
the 
ones of a bulk antiferromagnet at $x=\pm L/2$ [Fig.~\ref{Fig2}(a)]. 
The values $K_b$ and $v_b$ in the bulk region are choosen such that
Eq.~(\ref{eq:afm-lagrange}) correctly reproduces the dynamic susceptibility
of a 3D AF ordered state. From the nonlinear sigma 
model description~\cite{allen:97}, we estimate 
$v_b \simeq \sqrt{3} Ja/\hbar$ and $K_b \simeq 4 \sqrt{3}/\pi$.
The spin conductance then follows from 
$G = [(g \mu_B)^2/\pi \hbar] 
\lim_{\omega_n \rightarrow 0} \omega_n G_{\phi \phi}(x,x^\prime,\omega_n) $ 
where
the time ordered Green's function 
\begin{equation}
G_{\phi \phi}(x,x^\prime,\omega_n) = \int_0^{\infty} d 
\tau \, e^{-i \omega_n \tau}  \langle T_\tau  \phi(x,\tau) 
\phi(x^\prime,0)\rangle, 
\label{eq:afm-conductance}
\end{equation}
must be evaluated for the {\it inhomogeneous} system including the
transition regions~\cite{maslov:95}. For
given $x^\prime \in [-L/2,L/2]$, $G_{\phi \phi}(x,x^\prime,\omega_n)$ is 
obtained
from the ansatz $G_{\phi \phi} (x,x^\prime,\omega_n) 
= a \exp[\omega_n x/v(x)]
+ b \exp [-\omega_n x /v(x)]$ for the four regions $x<-L/2$, $-L/2<x<x^\prime$,
$x^\prime<x<L/2$, and $L/2<x$. The boundary conditions for the
spin current are automatically satisfied by evaluating  
Eq.~(\ref{eq:afm-conductance}). We find that $\lim_{\omega_n \rightarrow 0} 
\omega_n G_{\phi \phi}(x,x^\prime,\omega_n) = 1/2 K_b$ is 
independent of $x$, $x^\prime$  and of the 
parameters $K$ and $v$ of the spin chain. The spin conductance  at $T=0$,
\begin{equation}
G =  \frac{(g \mu_B)^2}{h K_b}
\label{eq:afm-conductance2}
\end{equation} 
depends only on the parameter $K_b$ of the bulk system. 

We next argue that Eq.~(\ref{eq:afm-conductance2}) remains valid also for
a finite temperature and an offset magnetic field. Both 
analytical~\cite{eggert:94} and recent numerical~\cite{castella:95} analysis 
indicate that, even at finite $T \ll J/k_B$, Eq.~(\ref{eq:afm-lagrange}) 
describes the low energy behavior of the spin-$1/2$ chain
correctly. Similarly, a finite offset magnetic field, $g \mu_B B \ll J$,  
suppresses quantum fluctuations of $\hat{s}_z$ in the bulk, leading only to a 
slight decrease of $K_b$. Our result 
Eq.~(\ref{eq:afm-conductance2}) is hence robust both for
finite temperatures and magnetic fields. In summary, an AF 
spin-$1/2$ chain suspended between AF reservoirs acts as a transport channel 
for magnetization with a spin conductance of order $(g \mu_B)^2/h$.
For $N$ parallel spin chains 
with vanishing interchain exchange interaction, each chain acts as 
an independent transmission channel and $G$
increases by a factor $N$. The spin conductance is quantized in 
units of order $(g \mu_B)^2/h$.

Preparation of a sample as shown in 
Fig.~\ref{Fig1}(a) is challenging. A promising strategy is the use of a 
bulk material with an intrachain exchange 
$J$ much stronger than the interchain exchange, such as 
Sr$_2$CuO$_3$. If heated to temperatures  $T$ much larger 
than the N{\'e}el ordering temperature $T_N$, the spin chains decouple and 
magnetization is transported predominantly 
along the spin chains. Hence, an AF wire heated to  $T>T_N$ in 
its  central part, but cooled to $T \ll T_N$ 
at its ends [Fig.~\ref{Fig2}(b)] provides a realization of the system 
in Fig.~\ref{Fig1}(a). Recent experiments~\cite{sologubenko:01} provide 
strong evidence that elementary excitations in various quasi-one-dimensional
systems have mean-free paths of several hundred nanometers at temperatures
up to $50$~K. The mean-free path is limited by the defect concentration in
the samples. For $L<1$~$\mu$m, transport through the system shown in 
Fig.~\ref{Fig2}(b) then is indeed ballistic as assumed above
[Eq.~(\ref{eq:afm-conductance2})].

\begin{figure}
\centerline{\mbox{\includegraphics[width=8cm]{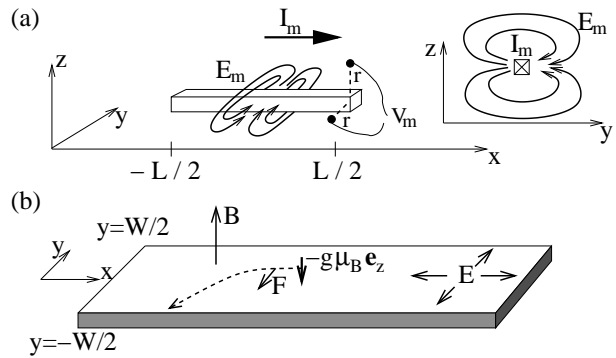}}}
\caption{
(a) A current of magnetic dipole moment $I_m$ produces an electric 
dipole field leading to a measurable voltage $V_m$. 
(b) Magnetic dipoles $-g \mu_B {\bf e}_z$ driven by a magnetic
field gradient $\nabla B$ in an inhomogeneous electric field 
${\bf E}({\bf x})$ 
experience a force ${\bf F}$ analogous to the Lorentz force. 
}
\label{Fig3}
\end{figure}

{\it Detection of spin currents. --} 
A current of magnetic dipoles produces 
an electric dipole field. The electric field is most easily calculated
by decomposing the magnetization current into contributions propagating at 
a certain velocity $v$, 
$I_m = g \mu_B \sum_v n(v) v$, where $n(v)$ is the line density of magnetic 
dipoles with velocity $v$. For each $v$, the electric field in the laboratory 
frame is obtained by a Lorentz transform of the magnetic dipole field 
in the comoving frame. Summing over $v$, we find that the total electric 
dipole field 
[Fig.~\ref{Fig3}(a)]
\begin{equation}
{\bf E}_{m}
({\bf x}) = \frac{\mu_0}{2\pi}\frac{I_m}{r^2} \left(
0, \cos 2 \phi, - \sin 2 \phi \right)
\label{eq:e-field}
\end{equation}
depends only on $I_m$. Here,
$\sin \phi = y/r$, $\cos \phi = z/r$, and $r =\sqrt{y^2 + z^2}$. For a 
numerical estimate, we now consider $N$ parallel 
uncoupled AF spin-1/2 chains connecting two AF reservoirs. 
With Eqs.~(\ref{eq:afm-conductance2}) and (\ref{eq:e-field}),
\begin{equation}
|{\bf E}_m({\bf x})| \sim  N   \frac{\mu_0}{2 \pi} \frac{(g \mu_B)^2}{h}
\frac{\Delta B}{r^2} 
= N \frac{g^2}{4} \times 10^{-19} \frac{\Delta B[T]}{r[m]^2} \frac{V}{m}.
\label{eq:e-field2}
\end{equation}
Even for moderate $\Delta B = 10^{-3}$~T and large  $r=10^{-5}$~m, the 
magnetization
current transported by $N \simeq 10^{4}$ parallel spin chains leads to an 
electric field $E_m \sim 10^{-8}$~V/m. 
The voltage drop between the two points $(0,r,0)$ and $(0,0,r)$ indicated in 
Fig.~\ref{Fig3}(a) is then $V_m = E_m r \simeq 10^{-13}$~V, which is within 
experimental reach~\cite{remark4,remark5}.

{\it Spin currents in electric fields. --} A moving magnetic dipole moment 
also interacts with an external electric field ${\bf E}({\bf x})$, leading to 
phenomena analogous to the Hall effect. A magnetic dipole 
$-g \mu_B {\bf e}_z$ moving in an electric 
field acquires an Aharonov-Casher phase~\cite{aharonov:84} and  the spin
Hamiltonian is modified to
\begin{eqnarray}
\hat{H}& =& \frac{J}{2}  \sum_{\langle ij \rangle} \left[
\hat{s}_{i}^+ \hat{s}_{j}^- e^{-i \theta_{ij}} 
+ \hat{s}_{i}^- \hat{s}_{j}^+ e^{i \theta_{ij}} +
2 \hat{s}_{i,z} \hat{s}_{j,z} \right] \nonumber \\ && 
\hspace*{1cm} + g \mu_B \sum_i B_i \hat{s}_{i,z},
\label{eq:hham2}
\end{eqnarray} 
where $\hat{s}_j ^{\pm} = \hat{s}_{j,x} \pm i \hat{s}_{j,y}$ and
$\theta_{ij} = g \mu_B \int_{{\bf x}_i}^{{\bf x}_j} d{\bf x} \cdot 
({\bf E} \times {\bf e}_z)/\hbar c^2$. 
Introducing magnon creation and annihilation operators, 
Eq.~(\ref{eq:hham2}) can be rewritten in terms of magnons
with single-magnon Hamiltonian $\hat{h}$. From Eq.~(\ref{eq:hham2}),
\begin{equation}
\hat{h} = \frac{|J|Sa^2}{\hbar^2} (\hat{\bf p} - g \mu_B {\bf E}\times 
{\bf e}_z/c^2)^2 + g \mu_B B.
\label{eq:magnonh}
\end{equation}
Here, we discuss only the classical motion of magnons propagating with 
velocity ${\bf v} = - v_x {\bf e}_x$ in a 2D system of finite width $W$ in
the $y$ direction [Fig.~\ref{Fig3}(b)], where $I_m = g\mu_B n v_x W$, and 
$n$ is the magnon density. 
From the equation of motion implied by
Eq.~(\ref{eq:magnonh}), one obtains the force acting on a magnon, 
${\bf F} = - g \mu_B \nabla [B - ({\bf v} \times {\bf E}) \cdot {\bf e}_z/
c^2]$. The second term accounts 
for the interaction with the electric field. We now focus 
on ${\bf E} = E^\prime (x,y,-2z)$ with 
$E^\prime = {\rm const}$. Then, the equation of motion of the magnons is 
formally identical to that of electrons
in a constant magnetic field. Magnons are deflected  into the 
${\bf e}_y$ direction  perpendicular to the transport direction ${\bf e}_x$.
Stationarity is reached when the magnon repulsion
equals the driving force along ${\bf e}_y$ due to the electric field.
Taking into account only dipolar forces between the magnons, in
the stationary state $B - v_x E^\prime y/c^2$ is
constant as function of $y$. The difference in magnetic fields 
$\Delta B = B(y=W/2)-B(y=-W/2)$ is
related to the magnetization current density by the spin
Hall conductance $G_H$,
\begin{equation}
\frac{I_m}{W} = - G_H \frac{\Delta B}{W} 
= -\frac{g \mu_B n c^2}{E^\prime} \frac{\Delta B}{W}.
\end{equation}
In the hydrodynamic regime, the drift velocity $v_x$ is determined by the
magnon scattering time $\tau$. At low temperatures, $\tau$ is limited by
impurities in the sample. For $\tau$ on the order of $10^2$ -- $10^3$~ns, 
as measured for yttrium iron garnet (YIG) 
at $1$--$4$ K (Ref.~\onlinecite{douglass:63}), 
$\partial_x B 
= 10^6$~T/m, $J = 200$ K~$k_B$, $S=1$, and $a = 1$~\AA, the drift velocity is
$v_x = 10$ -- $10^3$~m/s. A variation of electric field $\Delta E =
E(y=W/2) - E(y=-W/2) = 10^7$~V/m across the magnetic system then would lead
to $\Delta B = 10^{-3}$ -- $10^{-1}$~G resulting from
the spin Hall effect. Thus, the spin Hall conductance $G_H$ is within 
experimental reach.

{\it Acknowledgements. --} This work was supported by the
EU TMR network MOLNANOMAG, no. HPRN-CT-1999-00012, the Swiss NCCR 
Nanoscience, DARPA, and the Swiss NSF.
We gratefully acknowledge discussions with C.~Bruder, C.~Egues, H.-A.~Engel,
H.~Gassmann,  A.~Khaetskii, F.~Marquardt, and  P.~Recher.


\begin{thebibliography}{99}
\bibitem{remark2}For the clean system considered here, the magnon transmission
coefficient is unity.
\bibitem{remark3}The analysis including the result in 
Eq.~(\ref{eq:afm-conductance2}) remains valid also for anisotropic spin-$1/2$ 
chains which show Luttinger liquid behavior.
\end{thebibliography}
\end{document}